\documentclass[a4paper,fleqn]{cas-dc}

\UseRawInputEncoding
\usepackage[authoryear]{natbib}
\usepackage{subcaption}

\def\tsc#1{\csdef{#1}{\textsc{\lowercase{#1}}\xspace}}
\tsc{WGM}
\tsc{QE}
\begin{document}
\let\WriteBookmarks\relax
\def\floatpagepagefraction{1}
\def\textpagefraction{.001}

\shorttitle{An analysis on OpenMetBuoy-v2021 drifter data and Lagrangian trajectory simulations}    

\shortauthors{Moerman et al.}  

\title [mode = title]{An analysis on OpenMetBuoy-v2021 drifter in-situ data and Lagrangian trajectory simulations in the Agulhas Current System}  


\author[1,2]{Bente Moerman}[type=editor, orcid=0009-0000-6878-8744]
\author[1,3]{{{\O}}yvind Breivik}[type=editor, orcid=0000-0002-2900-8458]
\author[1]{Lars R. Hole}[type=editor, orcid=0000-0002-2246-9235]
\cormark[1]
\ead{lrh@met.no}
\author[1]{Gaute Hope}[type=editor, orcid=0000-0002-5653-1447]
\author[4]{Johnny A. Johannessen} [type=editor, orcid=0000-0001-7591-9714]
\author[5]{Jean Rabault}[type=editor, orcid=0000-0002-7244-6592]

\affiliation[1]{organization={Norwegian Meteorological Institute},
    addressline={All\'{e}gt. 70}, 
    city={5007 Bergen},
    country={Norway}}
\affiliation[2]{organization={Institute for Marine and Atmospheric Research Utrecht,   Utrecht University},
    addressline={Princetonplein 5}, 
    city={Utrecht},
    postcode={3584 CC}, 
    country={the Netherlands}}
\affiliation[3]{organization={Geophysical Institute, University of Bergen},
    addressline={All\'{e}gt. 70}, 
    city={5007 Bergen},
    country={Norway}}
\affiliation[4]{organization={Nansen Environmental and Remote Sensing Center},
    addressline={Jahnebakken 3}, 
    city={Bergen},
    postcode={N-5007}, 
    country={Norway}}
\affiliation[5]{organization={Norwegian Meteorological Institute},
    addressline={Henrik Mohns plass 1}, 
    city={0371 Oslo},
    country={Norway}}

%\credit{Formal analysis, Visualization, Writing – original draft, Writing – review \& editing}

\cortext[1]{Corresponding author}


\begin{abstract}
In order to perform a sensitivity analysis of Lagrangian trajectory models, Lagrangian trajectory simulations have been compared to six OpenMetBuoy-v2021 drifter trajectories in the Agulhas Current System (Jan-Mar 2023). 
Three different Lagrangian trajectory simulations have been assessed: (1) two offline Lagrangian tracking tools, OpenDrift and Parcels, (2) three Eulerian ocean surface current products, HYCOM, Mercator and Globcurrent, and (3) the addition of wind and/or wave forcing parameterizations.
The latter has also been evaluated by strong ocean current, high wind speed and Stokes drift regimes.

Firstly, using the same time stepping scheme and linear interpolation methods, the different Lagrangian simulators OpenDrift and Parcels, performed identically.
Secondly, the Globcurrent product showed the highest mean skill of the three ocean current products, although it underestimated the speed for strong ocean currents due to its spatial resolution.
The HYCOM and Mercator model simulations showed, respectively, 40\% and 15\% lower skill than the Globcurrent simulations.
Finally, the addition of the Stokes drift and a wind drift factor (WDF), improved the Lagrangian simulation performance in skill and speed, especially in high wind (>10 m/s) and/or Stokes drift regimes (>0.15 m/s).
The optimal WDF for the OpenMetBuoy-v2021 is found to be ~1.8\% and ~2.3\% for simulations including and excluding Stokes drift forcing respectively. 
To further improve the incorporation of Stokes drift and direct wind drag on the trajectory simulations, a more physically based solution is advised as there are still numerous wind and wave related processes that remain unresolved, like wave-current interactions and vertical shear.

To statistically strengthen the conclusions from this research, incorporating additional observed drifter trajectories would be highly favourable.
\end{abstract}


% \begin{highlights}
% \item 
% \item 
% \item 
% \end{highlights}

% Keywords
% \begin{keywords}
%  \sep Lagrangian trajectory modelling \sep OpenMetBuoy-v2021 \sep Wind drift factor \sep Agulhas Current \sep OpenDrift \sep Parcels \sep Ocean Surface Current \sep
% \end{keywords}

\maketitle

% Main text
\section{Introduction}\label{Introduction}
Human activities in the oceans predominantly take place at the ocean surface. 
To enhance marine safety, an adequate level of detail from observations and a sufficient predictive skill of numerical models of the ocean surface current is required (\cite{Rohrs2023}).
Although, the application of ocean currents from operational forecasting has long been hindered by the chaotic nature of oceanic flow and the limited coverage of observations, recently, high-resolution ocean models are enhancing in predictive skill (\cite{Christensen2018}). \\
\\
One application of the predicted ocean surface current by operational ocean circulation models, is Lagrangian trajectory simulations.
Lagrangian trajectory simulations make use of the Eulerian ocean current to simulate the movement of a particle, object or substance at the ocean surface. 
Research on the drift at the surface of the ocean with Lagrangian particle tracking tools has numerous academical and practical purposes including the faith of pollutants in the ocean as marine plastics (\cite{VanSebille2015}), search and rescue (\cite{Breivik2013}), the spread of living organisms (fish eggs, larvae, plankton blooms, etc.) (\cite{Rohrs2014}), oil spill evolution and clean-up (\cite{Mellor2016}) and hydrodynamic connectivity (\cite{VanSebille2010}). 
All have a large environmental and societal impact and require quick action.
Hence, acquiring a thorough understanding of the limitations and optimal utilization of Lagrangian trajectory simulations is highly advantageous.\\
\\
The Eulerian ocean models, used for Lagrangian trajectory simulations, are lacking proper incorporation of dominant processes and dynamics in the upper ocean. 
For example, Eulerian ocean models do not resolve the wind-induced shear stress gradient in the ocean surface layer (e.g. \cite{Chassignet2006}; \cite{Dagestad2019}). 
In addition, incomplete treatment of coupling mechanisms between ocean and atmosphere, result in unresolved air-wave-sea interactions. 
Especially during extreme weather events, air-wave-sea interactions have a large impact on the surface drift (\cite{Rohrs2023}).
Wind and wave parameterizations are often incorporated in Lagrangian simulations to overcome the unresolved surface processes in the Eulerian ocean models. 
However, it remains difficult to correctly describe the combination of current, wind and wave driven transport in trajectory simulations (\cite{Rohrs2023}). 
The research addressed in this paper is therefore motivated by the need to advance the quantative  understanding of the limiting factors in the Lagrangian trajectory simulation performance related to Eulerian ocean surface current products and wind and wave parameterizations. 
The data and methods are addressed in Section 2, followed by the presentation of the results in Section 3. 
The discussion and conclusion are then provided in Section 4.

\section{Study area, Data and Methods}\label{Methods}
\subsection{Study Area}
The Agulhas Current, the western boundary current of the South West Indian Ocean gyre, is transporting warm and saline water southwestwards into the South Atlantic. 
The current is one of the strongest ocean surface currents and is characterized by intense current shears and horizontal temperature gradients. 
It is highly nonlinear and has a  large impact on the regional marine climate, the regional weather and the global climate (\cite{Lutjeharms2006}, \cite{Beal2011}). 
The mesoscale to sub-mesoscale ocean processes and dynamics are complex and occur across a broad range of temporal and spatial scales (\cite{Krug2014}, (\cite{Hart-Davis2018}). 
Moreover, the greater Agulhas Current regime can be divided in three main parts by their characteristic dynamics, namely the Agulhas Current core, the Agulhas Retroflection and the Agulhas Return Current (\cite{Lutjeharms2006}). \\
\\
The Agulhas region is subject to various wind regimes.
The predominant wind direction is Southwest, but Northeasterly winds also occur frequently when the south Indian high pressure cell is situated east of South Africa (\cite{Barnes2020}). 
The wave climate in the Agulhas Current System is dominated by midlatitude cyclones in the Southern Ocean that generate long-period southwesterly swell (\cite{Rautenbach2020}). 
Occasionally, wave signals from the Indian Ocean reach the east coast of South Africa due to passing tropical cyclones and slow-moving low-pressure systems (\cite{Fitchett2014}). \\
\\
In January 2023 six wave drifters (OpenMetBuoy-v2021, OMB, \cite{Rabault2022}) were deployed in the Agulhas Current System by the Norwegian Meteorological Institute during the One Ocean Expedition (\url{https://oneoceanexpedition.com}), a global ocean circumnavigation onboard of the Norwegian tall ship, Statsraad Lehmkuhl.
A sensitivity analysis is carried out by which Lagrangian trajectory simulations are compared to the observed OMB data.
Three aspects are considered. 
First, the performances of two Lagrangian models, OpenDrift (\cite{Dagestad2018}) and Parcels (\cite{Lange2017}; \cite{Delandmeter2019}) are assessed.
Second, the performances of three Eulerian ocean surface current products, notably, HYCOM (\cite{Bleck2001}; \cite{Halliwell2004}; \cite{Chassignet2007}) and Mercator (\cite{GASPARIN2018}) and the observation-based Globcurrent field (\cite{Rio2014}) are examined.
Third, the addition of wind drag parameterization and/or wave forcing are evaluated. 
As the wind drag has a strong regional and seasonal dependency (\cite{Rio2003}), the  
addition of wind and wave parameterizations are also evaluated for strong ocean current, high wind speed and Stokes drift conditions.

\subsection{Lagrangian trajectory models}
Drifter simulations are conducted with two open-source, offline particle tracking codes, OpenDrift (\cite{Dagestad2018}), under development by the Norwegian Meteorological Institute, and Parcels (\cite{Lange2017}; \cite{Delandmeter2019}), under development at the Institute for Marine and Atmospheric Research at the University of Utrecht. \\
\\
To calculate the trajectory of a particle, the Lagrangian trajectory simulations update the position of the particle by integration of the velocity field over one time step: 
\begin{equation} \label{eq:int}
    X(t+\Delta t) = X(t) + \int^{t+\Delta t}_{t} v(x(\tau),\tau)d\tau,
\end{equation}
where $v(x,\tau)$ is the Eulerian velocity field and can be converted to the Lagrangian velocity $dX(t)/dt$ at $X(t)=x$. 
Different Lagrangian tracking tools are invoked by the implementation of an additional term on the right hand side of the equation that compensates for unresolved physics in the Eulerian ocean model. 
Besides, there are multiple time-stepping schemes that can be used to compute the integration. 

The accuracy of the Lagrangian trajectory computation is mainly dependent on (1) the accuracy of the time stepping scheme and (2) the accuracy of the used interpolation scheme used to estimate the velocity of the particle from the Eulerian velocity field (\cite{VanSebille2018}).

The time-stepping scheme used in this analysis is the Euler forward method, as it is the default scheme in OpenDrift and it has a shorter computation time then other, higher order methods.
For the interpolation OpenDrift uses bilinear interpolation (\textit{scipy.ndimage.-map.coordinates}) as default method. 
OpenDrift may also extrapolate data towards land if there is an area without data between the ocean pixel at the edge of the ocean model and the considered coastline determined by a separate land mask (\cite{Dagestad2018}). 
On the other hand, Parcels has multiple interpolation options based on the grid of the input data from the ocean model. 
There are interpolation schemes available for A-, B-, and C-based staggered grids, rectilinear and curvilinear horizontal meshes, and vertical levels determined by \textit{z} and \textit{s}. 
For an A grid, Parcels uses linear interpolation. (\cite{Delandmeter2019}) 
Only linear interpolation has been used in this analyses.

The specifications of the Lagrangian tracking tools OpenDrift and Parcels are summarized in Table \ref{tab:lag_models}.

\begin{table*}[]
    \centering
    \begin{tabular}{m{2.5cm} m{6.6cm} m{6.8cm}}
        \hline
        & OpenDrift & Parcels \\
        \hline
        Website & \url{https://opendrift.github.io/index.html} & \url{https://docs.oceanparcels.org/en/latest/index.html} \\
        License & GPLv2 & MIT \\
        Key citation & \cite{Dagestad2018} & \cite{Lange2017} \\
        OGCMs supported & ROMS & NEMO, OFES, Globurrent, MOM, ROMS\\
        Language(s) & Python & Python user interface, autogenerated C\\
        Primary use & Oil drift, search-and-rescue, plankton drift, drift of toxic chemicals & Large scale oceanography; Individual based Modelling; Teaching; Plastic \\
        Advection method & Explicit Euler, RK2, RK4 & RK4, RK45, Explicit Euler; Analytical; Extensible interface for custom advection methods \\
        Diffusion method & Obtain data from ensemble models & Extensible interface for Random Walk and custom behaviour \\
        Grids supported & Generic, but dedicated readers must provide data on regular projected grids for further internal calculation & Arawaka A, B and C; unstructured meshes planned \\
        Key strengths & Open-source, customizable, extendable & Easy to use; customizable, extension interface and automated performane optimization \\
        Shortcomings & e.g. the number of elements must be determined before simulation starts & Support for unstructured grids in progress \\
        \hline
    \end{tabular}
    \caption{Overview of the specifications of Lagrangian trajectory models OpenDrift and Parcels.}
    \label{tab:lag_models}
\end{table*}

\subsection{Meteorological and Ocean forcing} \label{forcing}
Eulerian ocean models differ from each other in meshes, grids and vertical coordinates, but also in approximations, (un)resolved processes, parameterizations, coupling and complexity.
All above affects the model performance. 
For instance, every grid type has their own strengths and weaknesses, which makes the choice of grid a trade-off between numerical noise and inaccuracy (\cite{Marsh2021b}). 

In this research three Eulerian ocean products will be considered. 
Two model-based products, the HYbrid-Coordinate Ocean Model (HYCOM), and the Mercator Surface Merged Ocean Currents (SMOC) from the Copernicus Marine Environmental Service (CMEMS) and the observation-based Globcurrent product (\cite{Rio2014}) are analysed. 
In Table \ref{tab:ocean}, an overview of the specifications of the three ocean current products can be found.
The Mercator product delivers four different currents, a Total current and separated into three parts, the Navier-Stokes current (NS), Stokes drift and Tidal current. 
For this study Mercator total and Mercator NS will be considered.
The Globcurrent product delivers a global total surface current from the geostrophic current, obtained with satellite altimetry, combined with modeled Ekman current processing.

\begin{table*}
    \centering
    \begin{tabular}{ p{2.8cm}  p{4.2cm}  p{4.2cm}  p{4.2cm} }
         \hline 
         % \multirow{2}{*}{Name} & \multicolumn{2}{|c|}{Model based + assimilation} & \multicolumn{1}{|c|}{Observation based} \\
         & HYCOM & Mercator (SMOC) & Globcurrent NRT \\ 
         \hline %\hline
         Product type & Model based & Model based & Observation based: altimeter geostrophy + Ekman model \\ %\hline
         System & & GLO12v4 & MULTIOBS GLO PHY NRT 015 003 \\ %\hline
         Temporal frequency & 3h & 1h & 6h / 1d / 1m \\ %\hline
         Spatial resolution & 1/12\textdegree lon x 1/24\textdegree lat & 1/12\textdegree & 1/4\textdegree  \\ %\hline
         Grid & Standard grid & Standard grid (interpolated from native C) & Standard grid\\ %\hline
         Vertical resolution & Hybrid & 50 levels - Arawaka native & 0 (surface) and -15 m\\ %\hline
         Vertical coordinate & Hybrid: isopycnic ($\rho$), fixed depth (z) or pressure (p) or terrain-following levels & fixed depth (z) & \\ %\hline
         Atmospheric forcing & 0.5\textdegree NAVGEM & 0.125\textdegree ECMWF Atmospheric forecasting model & ERA5*  \\ %\hline
         Data assimilation & NCODA (Cummings, 2005; Cummings and Smedstad, 2013) & SAM2 &  \\ %\hline
         Originating centre & Naval Research Laboratory: Ocean Dynamics and Prediction Branch & Mercator Océan International & CLS (France)\\
         \hline 
    \end{tabular}
    \caption{Specifications of the Eulerian ocean surface current products.}
    \label{tab:ocean}
\end{table*}

The atmospheric forcing is provided by the wind forecast of the European Centre for Medium-Range Weather Forecasts (ECMWF).
The spatial resolution of the product is around 25 km kilometres. 
The time step of the model is 1 hour and the data is updated twice a day. 
The 10-meter wind speed in meridional and zonal directions, were extracted at a 3-hour temporal scale. 

The wave forcing in this study is provided by ECMWF Ocean Wave Model (ECWAM, \cite{Ardhuin2010}; \cite{Janssen2014}).
ECWAM is a global model and coupled to an atmospheric forecast model and the dynamical ocean model NEMO. 
ECWAM assimilates space-borne altimeter significant wave height (SWH) data.
The spatial resolution is tens of kilometres. 
The time step of the model is 1 hour and the data is updated twice a day. 
Six-hourly data of the Stokes drift (meridional and zonal direction) and SWH have been extracted.

\subsection{The OpenMetBuoy-v2021}
\begin{figure}
    % \centering
    \begin{subfigure}{0.23\textwidth}
    \includegraphics[width=\linewidth]{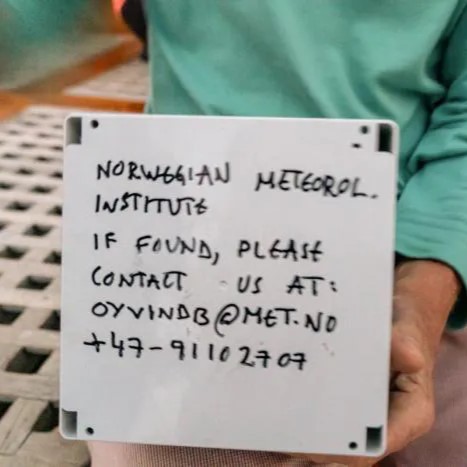}
    \caption{}
    \label{fig:OMB}
    \end{subfigure}
% \end{figure}
    \hfill
% \begin{figure}
    % \centering
    \begin{subfigure}{0.23\textwidth}
    \includegraphics[width=\linewidth]{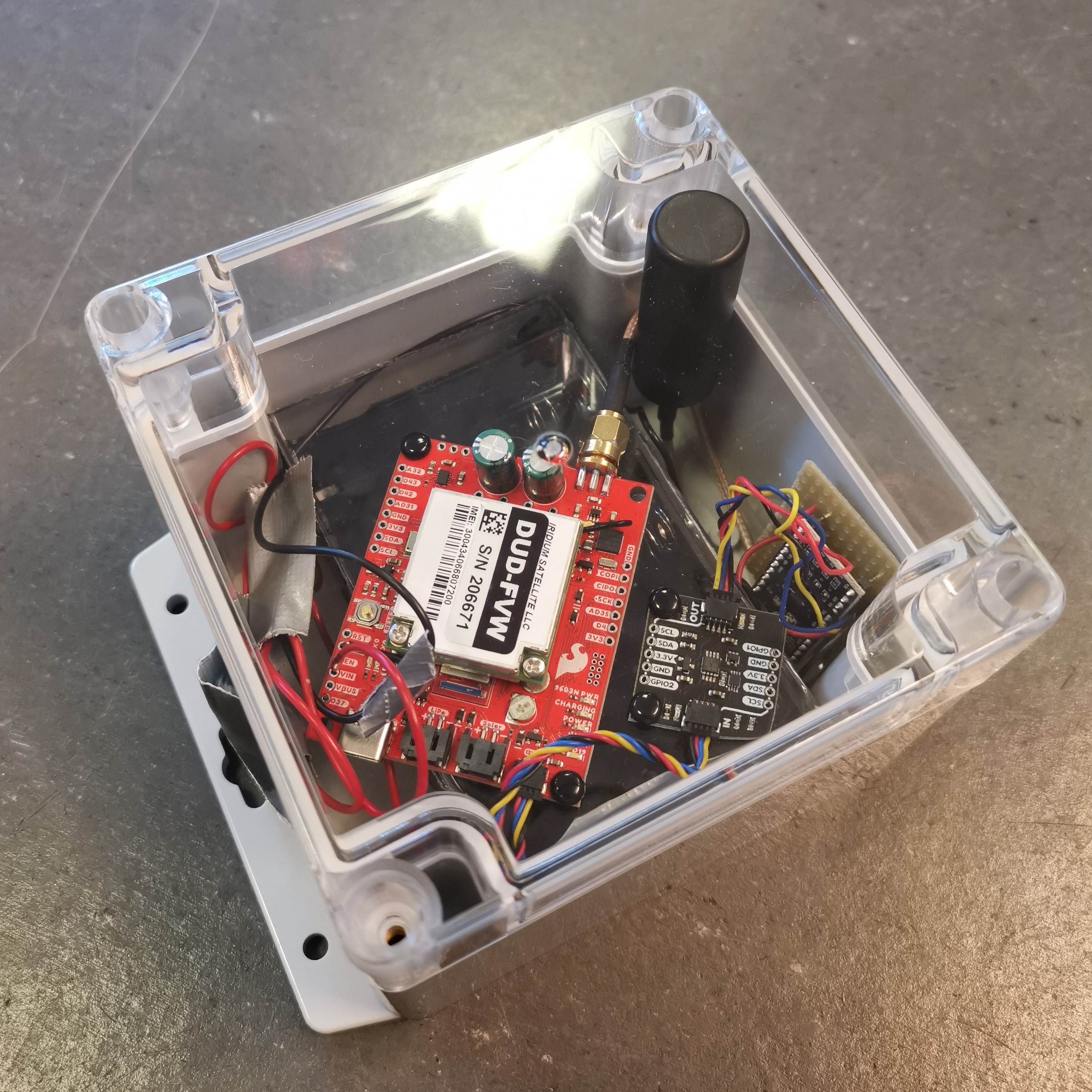}
    \caption{}
    \label{fig:OMB_b}
    \end{subfigure}
    \caption{(a) The OMB on board of the Statsraad Lehmkuhl during the OneOcean expedition before deployment in the Agulhas Current System, January 2023 (Photo: Susanne Nj\o lstad Skandsen). (b) A fully assembled buoy with a box size of 12x12x9 cm, including 3 D-size battery holders (empty on the picture), all the components, and the 9dof sensor (\url{https://github.com/jerabaul29/OpenMetBuoy-v2021a/tree/main}, accessed on 14 December 2023).}
\end{figure}

The OpenMetBuoy-v2021 (OMB), developed by \cite{Rabault2022} at the Norwegian Meteorological Institute (Figure \ref{fig:OMB}), is designed as a cost-effective, easy to build and open source drifter to measure the ocean surface drift and monitor waves in the open ocean and in sea ice. 
The data acquisition and processing are both performed on a high-performance microcontroller, which makes the instrument power efficient. 
The geographical positioning of the buoy is obtained with a GNSS module and is sampled every 30 minutes. 
The communication of the instrument goes via iridium to enable global coverage. 
The drifter has a box shape of size 12 cm x 12 cm x 9 cm, and weighs between 0.3-0.5 kg when two lithium D-cell batteries are used. 
The design and signal processing code of the OMB are validated through test campaigns in the open sea, on sea ice and in the laboratory, see \cite{Rabault2022,rabault2023dataset}. 
The wave spectra of the OMB were found to be in agreement within 5\% with the commercial Spotter buoy by SOFAR (https://www.sofarocean.com/products/spotter, accessed in December 2023).
Based on these validation tests, the buoy is considered to be a robust and well-functioning drifter.

\subsection{The Deployment} 
The drifters used in this research were deployed by the Norwegian Meteorological Institute during the One Ocean Expedition (\url{https://oneoceanexpedition.com}).
The Norwegian tall ship, Statsraad Lehmkuhl, sailed 55,000 nautical miles around the world on which it continuously collected high-quality data for ocean research, in the fields of biology, chemistry and physics, with the goal to increase knowledge about the ocean and its role in sustainable development (\cite{Huse2023}). 
All the data collected during the expedition is available open source (\cite{OneOceanData}). 

\begin{figure*}
    \centering
    \includegraphics[width=\linewidth]{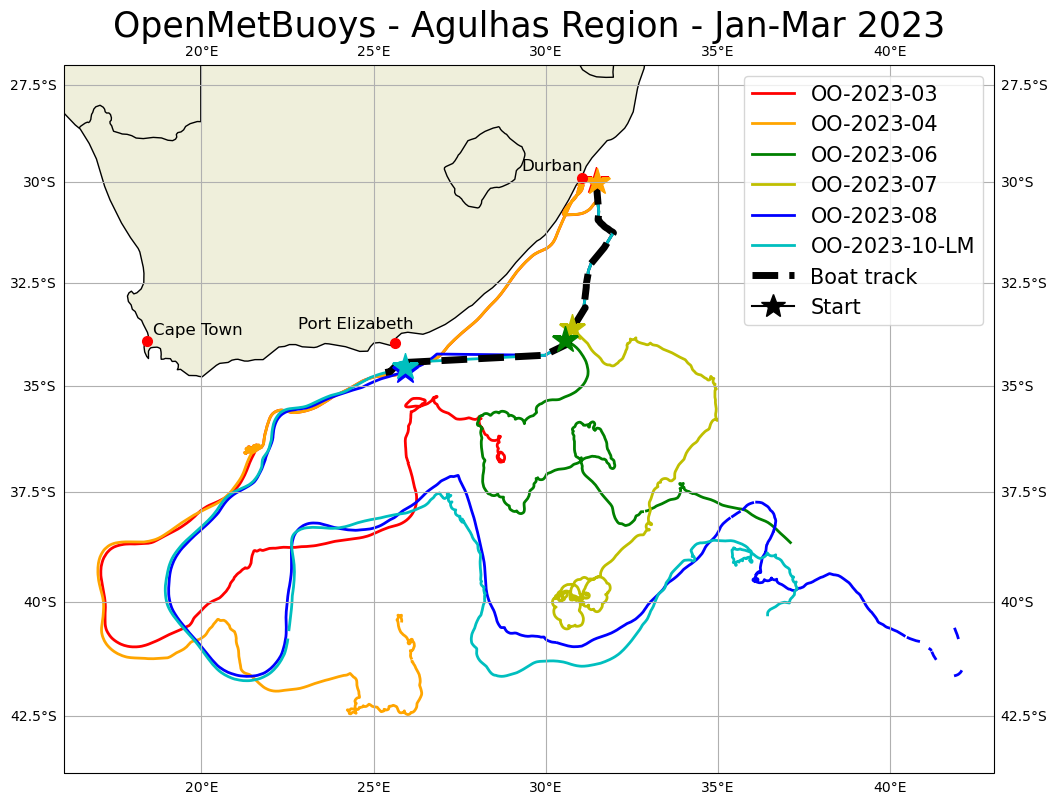}
    \caption{The OpenMetBuoy drifter trajectories in the Greater Agulhas Current System from January to March 2023. The deployment locations are indicated with a star. Drifter OO-2023-08 lost connection sometimes during the last days of the experiment and therefore shows a noncontinuous trajectory.}
    \label{fig:drifters}
\end{figure*}

The OMBs were deployed from the tall ship on its leg from Maputo, Mozambique to Cape Town, South Africa. 
The drifters were deployed in three different locations in the Agulhas Current System. 
One pair was deployed in the core of the Agulhas Current around 100 km off the coast at Durban.
Two drifter were deployed about 50-100 km apart, one in a clockwise, and one in an anticlockwise eddy between the Agulhas Current and the Agulhas Return Current.
Another drifter pair was deployed further South in the Agulhas Current around 100 km off the coast from Port Elizabeth. 
All the drifters transmitted data for approximately two months from about the first week of January to the second week of March.   
The six drifter trajectories are displayed in Figure \ref{fig:drifters} accompanied by corresponding drifter specifications, deployment sites, drift duration, environmental conditions and drifter speed in Table \ref{tab:drifters}.

\begin{table*}[]
    \centering
    \begin{tabular}{m{2.5cm} m{2.5cm} m{2.3cm} m{2.0cm} m{2.0cm} m{2.0cm}}
        \hline
        Name/Nr. & Deployment location & Time & Hs (m) mean/max. & Trajectory length (km) & Drifter speed (m/s) mean/max. \\
        \hline 
        OO-2023-03 & Durban & 5 jan - 9 mar & 2.7/7.4 & 4788 & 0.85/5.82 \\
        OO-2023-04 & Durban & 5 jan - 8 mar & 3.0/8.4 & 4518 & 0.81/6.74 \\
        OO-2023-06 & CCW Eddy & 8 jan - 9 mar & 2.7/7.1 & 3046 & 0.96/4.78 \\
        OO-2023-07 & CW Eddy & 8 jan - 9 mar & 3.0/7.8 & 3349 & 0.57/5.82 \\
        OO-2023-08 & Port Elizabeth & 9 jan - 4 mar & 3.0/9.1 & 5923 & 1.17/5.81 \\
        OO-2023-10-LM & Port Elizabeth & 9 jan - 10 mar & 2.9/9.1 & 5596 & 1.00/5.81 \\
        \hline
    \end{tabular}
    \caption{Overview of the OMB drift experiments in 2023, with the significant wave height (Hs) measured by the OMB and the wind and current data along the drifter trajectories extracted from the models. }
    \label{tab:drifters}
\end{table*}

\subsection{Experimental Set-Up}
The simulations are run in OpenDrift and Parcels using a Euler forward propagation scheme with a time step of 15 minutes and hourly output.  
The temporal and spatial scales of the drift simulations are based on the application of Lagrangian trajectory simulations for search and rescue missions and oil spill modelling (e.g. \cite{Parn2023}; \cite{Dagestad2019}).
Every 12h a new simulation is initiated from the observed OMB location at that time, with a simulation length of 24h.
To compare the observed drifter trajectories with the simulated trajectories, the drifter trajectories are cut into the same temporal segments of 24h. 
This adds up to a total amount of 707 segments/simulations per experiment. \\
\\
The following model is used for the drift velocity: 
\begin{equation}
    v = v_{surface} + v_{Stokes} + WDF*U_{10}
\end{equation}
The surface current ($v_{surface}$) is delivered by the Eulerian ocean forcing model.
If added, the surface Stokes drift ($v_{Stokes}$) and the 10-m wind speed ($U_{10}$) are provided by ECMWF (see section \ref{forcing}). 
The last term in the equation is the wind drag term. 
The wind drag is defined by the 10-wind speed multiplied by a wind drift factor (WDF).
To obtain the optimal WDF for the OMB, simulations will be conducted with 25 different wind drift factor's ranging in constant intervals from 0\% to 5\%. 
Their performance are analysed by comparison to the observed drifter trajectories by use of the python trajectory analysis package trajan (\url{https://opendrift.github.io/trajan/index.html}).

\begin{table*}[]
    \centering
    \begin{tabular}{|c c|c|c|c|c|c|}
         \hline
         & & Lagrangian model & Ocean forcing & Wind forcing & Wave forcing & Skill metrices  \\ \hline \hline
         Lagrangian model & 1 & Parcels & Globcurrent & - & - & MCSD, LWS, speed \\
         & 2 & OpenDrift & Globcurrent & - & - & MCSD, LWS, speed \\ \hline
         Ocean forcing & 2 & OpenDrift & Globcurrent & - & - & MCSD, LWS, speed  \\
         & 3 & OpenDrift  & HYCOM & - & - & MCSD, LWS, speed  \\
         & 4 & OpenDrift  & Mercator Total & - & - & MCSD, LWS, speed  \\
         & 5 & OpenDrift  & Mercator NS & - & - & MCSD, LWS, speed  \\ \hline
         Additional forcing & 5 & OpenDrift & Mercator NS & - & - & MCSD, LWS, WMS, speed  \\
         & 6 & OpenDrift & Mercator NS & ECMWF & - & MCSD, LWS, WMS, speed \\
         & 7 & OpenDrift & Mercator NS & - & ECWAM & MCSD, LWS, WMS, speed \\
         & 8 & OpenDrift & Mercator NS & ECMWF & ECWAM & MCSD, LWS, WMS, speed \\ \hline
    \end{tabular}
    \caption{An overview of the simulation experiments conducted in this research.}
    \label{tab:sims}
\end{table*}

\begin{figure*}
    \centering
    \includegraphics[width=\linewidth]{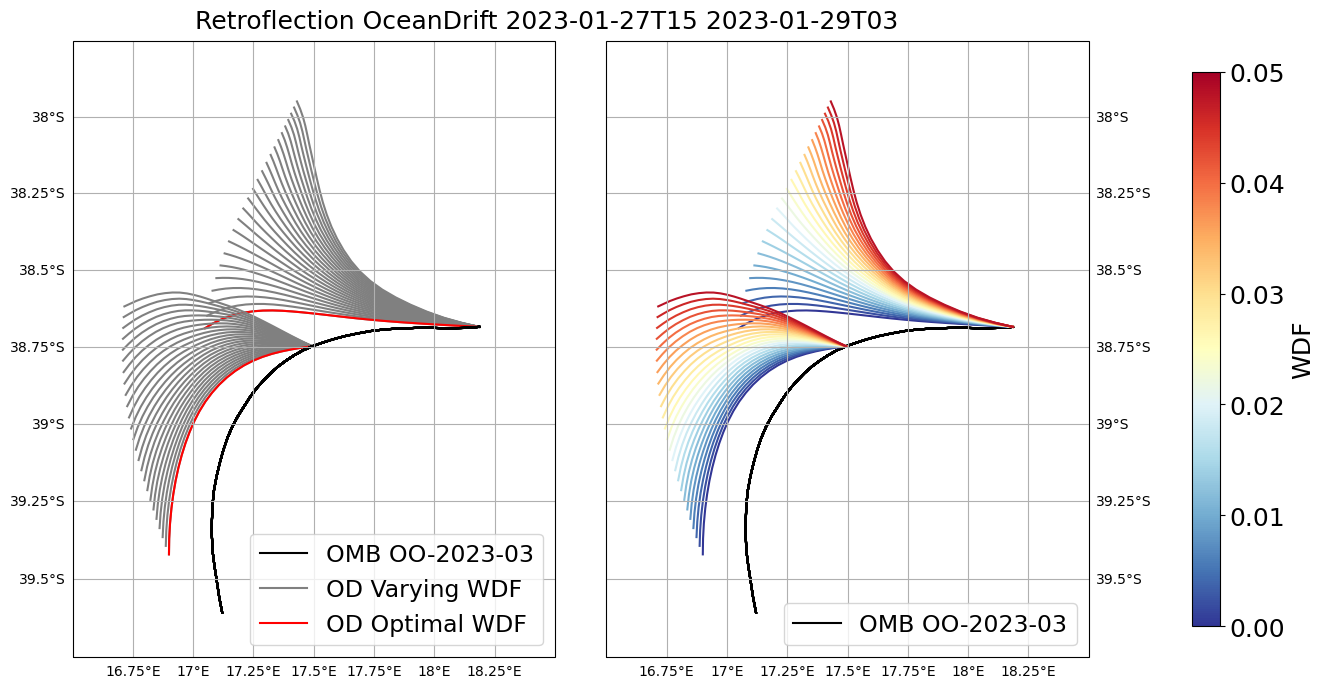}
    \caption{Illustration of the simulation set-up. Trajectory of OO-2023-03 (black) in the Agulhas Retroflection loop and simulated 24h trajectories, reinitialised every 12h from the location of the observed drifter trajectory. Simulations are run with ocean, wind and wave forcing (exp. 8 table \ref{tab:sims}). Right: Simulations with varying WDF from 0\% to 5\%. Left: Simulations with varying WDF (grey), indicating the simulation showing highest skill (red).}
    \label{fig:wdf}
\end{figure*}

Different factors in Lagrangian trajectory simulations are evaluated as indicated in Table \ref{tab:sims}: [1] the choice of Lagrangian model (Exp. 1-2), [2] the choice of Eulerian ocean forcing model (Exp. 2-5) and [3] the addition of wind and Stokes drift forcing (Exp. 5-8). 
Hence, multiple simulation experiments have been conducted.
In the ocean model analysis, the Mercator (NS and Total) simulations showed a significantly higher performance than the HYCOM simulations.
Mercator NS is therefore used for the additional forcing experiments. 
For the experiments including wind forcing, 25 simulations are conducted with different WDF ranging in constant intervals from 0-5\%, shown in the right Figure \ref{fig:wdf}.
The 25 simulations are compared to the observed drifter trajectory segments.
The simulation with the highest LWS is used for comparison with the other experiments, indicated by the simulation in red in Figure \ref{fig:wdf} (left). 
As well, the optimal WDF was analysed to find the best fitting WDF for the OMB.
The additional forcing experiments, are analysed as a whole, but also for large current and storm conditions, namely ocean currents > 1.25 m/s, wind speed > 10 m/s and Stokes drift > 0.15 m/s. 
The optimal WDF will also be analysed for the variant conditions. 

\subsubsection{Skill metrices}
In order to evaluate the performance of the simulated trajectories, various metrices are calculated: the Mean Cumulative Separation Distance (MCSD), the Liu-Weisberg skill score (LWS) and the Willmott skill score (WMS). \\
\\
\textbf{Mean Cumulative Separation Distance (MCSD).}
The MCSD is often used in trajectory analysis (for instance \cite{Haza2019}; \cite{Mheen2020}).
The MCSD takes into account separation distance between observed and simulated drifter on every time step and is calculated as:
\begin{equation} \label{eq:mcsd}
    MCSD = \frac{1}{T} \sum^{T-1}_{t=0} |x_i (t) - x_{obs}(t)|
\end{equation}
where $x_i(t)$ is the location of the simulated drifter $i$ at time $t$, $x_{obs}(t)$ is the observed drifter location at time $t$ and $T$ is the number of time steps in the drifter trajectory. \\
\\
\textbf{Liu-Weisberg skill score (LWS).} 
This metric is proposed by \cite{Liu2011}.
It accounts for the normalized cumulative separation distance (d), as well as the cumulative length of the trajectory (dl). 
By including the length of the trajectory into the skill score, this non-dimensional skill score indicates the relative performance of the simulations in modelling both strong currents and weaker currents (\cite{Liu2014}). \\
\\
The skill score is based on the ratio between d and dl: 
\begin{equation}
    s = \frac{\sum_{i=1}^{N} d_i}{\sum_{n=i}^{N} dl_i}
\end{equation}
where $d_i$ is the separation distance between the observed and the modelled locations of the drifters at time step i after the initialization of the model, $dl_i$ is the cumulative length of the observed drifter trajectory and N is the total number of time steps.
The LWS is calculated as follows: 
\begin{equation}\label{eq:liuweis}
    LWS = 
        \begin{cases}
        1 - \frac{s}{n} , (s \leq n) \\
        0 , (s > n), 
        \end{cases}
\end{equation}
where $n$ is the tolerance threshold, which is set to $n=1$ in this research.
The choice of this threshold is arbitrary, however it defines the requirements/expectations to the model (\cite{Liu2014}). 
A higher (lower) threshold should be used when there is lower (higher) expectations for the performance of the model. 
The LWS ranges from 0 to 1, where $LWS = 1$, indicates perfect alignment of the observed and simulated trajectories and if $LWS = 0$ there is no alignment at all. \\
\\
\textbf{Willmott skill score (WMS).} 
This index is proposed by \cite{Willmott1981} and is calculated by latitude and longitude separately.
This creates the ability to obtain differences in skill in these two directions. 
Moreover, this skill score is considered the most appropriate skill metric for evaluation of hydrodynamic models as it considers both the type and magnitude of potential correlations (\cite{Allen2013}; \cite{Willmott2012}). \\
\\
The WMS is defined by: 
\begin{equation} \label{willmott}
    WMS = 1 - \frac{\sum_{i=1}^N (P^i - O^i)^2}{\sum_{i=1}^N (|P_i^*| + |O_i^*|)^2}
\end{equation}
where P is positional time series (longitude or latitude) of the simulated trajectories and O is the corresponding positional time series of the observed trajectory.
$P^*$ and $O^*$ are defined as $P^* = (P - \overline{O})$ and $O^* = (O - \overline{O})$ , where $\overline{O}$ is the mean of the time series. 
$d$ is a measure for the degree to which the  deviations between the observed trajectory and the mean observed location ($O_i - \overline{O}$) correspond to the deviations between the simulated trajectory and the mean observed location ($P_i - \overline{O}$) in magnitude and sign. 
As the LWS, the WMS ranges from 0 to 1, where 1 represents perfect agreement of the simulated and observed trajectories and 0 indicates total disagreement. \\
\\
\textbf{Improvement Factor (IF).} 
Additionally, to compare the performance of the simulations run with only ocean forcing to the performance of simulations run with additional velocity field(s) and the performance of the ocean models HYCOM and Mercator with the observation based Globcurrent, an improvement factor is computed as the percentage decrease (increase) of the MCSD (LWS), following \cite{Haza2019}: 
\begin{equation} \label{eq:IF}
    IF_{MCSD} = (\frac{\sum MCSD_{ocean}}{\sum MCSD_{+wind/+wave/all}} - 1)* 100\% 
\end{equation}
\begin{equation}
    IF_{LWS} =  \frac{\sum LWS_{+wind/+wave/all} - \sum LWS_{ocean}}{\sum LWS_{ocean}} * 100\%
\end{equation}

\section{Results}\label{Results}
\subsection{Lagrangian models: OpenDrift \& Parcels}
The simulations done with the two models show fairly identical results in MCSD, LWS and speed compared to the observed drifter trajectories. 
Differences in speed between the OpenDrift and Parcels simulations are largest for the trajectory segments with the largest observed drifter speeds. 
However, the correlation coefficient between the Parcels and OpenDrift simulated drifter speeds is still 1.
The slight differences in speed are probably due to the small differences between the interpolation methods of the Lagrangian trajectory simulation models, though the differences are minimal. 

\subsection{Ocean forcing} \label{sec:results_ocean_forcing}
\begin{figure*}
    \centering
    \includegraphics[width=\linewidth]{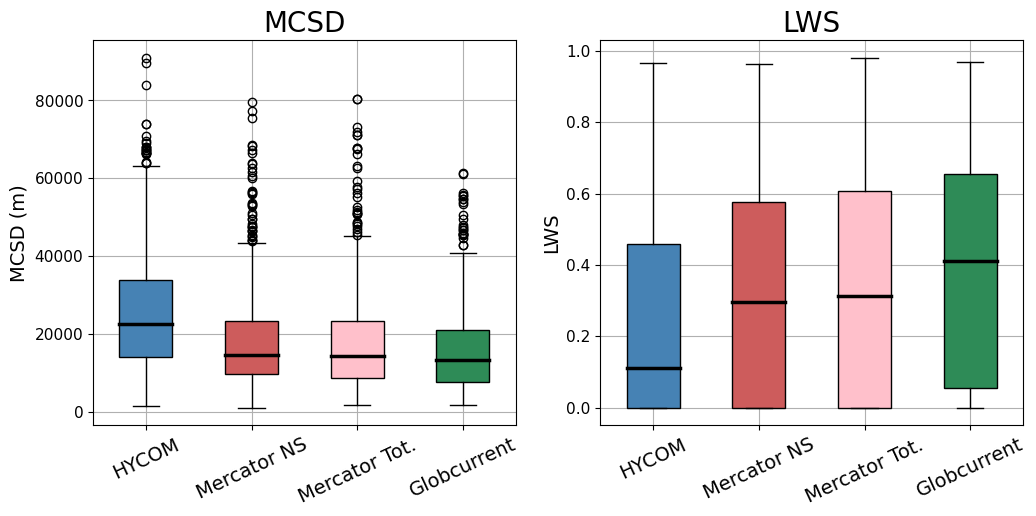}
    \caption{Box and whisker plots for the MCSD (Eq.\ref{eq:mcsd}) and LWS (Eq. \ref{eq:liuweis}) between the six observed drifter trajectories and the OpenDrift simulations forced by four different ocean current products. The box extends from the lower to upper quartile of the data and the horizontal line within the box represents the median value. The whiskers are plot from the lower quartile of data minus 1,5 times the IQR to the upper quartile plus 1,5 times the IQR. Flier points are plotted for data points outside the whisker range.}
    \label{fig:ocean_models}
\end{figure*}

In this section the OpenDrift simulations with four different global ocean surface current products are analysed, namely HYCOM, Mercator Total Current, Mercator Navier-Stokes (NS) and the Globcurrent product. 
In figure \ref{fig:ocean_models} the MCSD and the LWS between the simulated and the observed drifter trajectories are shown. 
The simulations ran with the HYCOM model show significantly larger MCSD than the other two models with a mean of 25.6 km compared to 18.4 km, 17.7 km and 15.9 km for Mercator NS, Mercator total and Globcurrent respectively. 
For the LWS, a similar pattern is observed with mean values of 0.23, 0.33, 0.35 and 0.39 for HYCOM, Mercator NS, Mercator total and Globcurrent respectively. 
As the Mercator NS and Total Current components show fairly similar results in MCSD and LWS and a separate Stokes drift field will be added to the simulations, the Mercator NS product is considered for the continuing of this research.  

\begin{figure*}
    \centering
    \includegraphics[width=\linewidth]{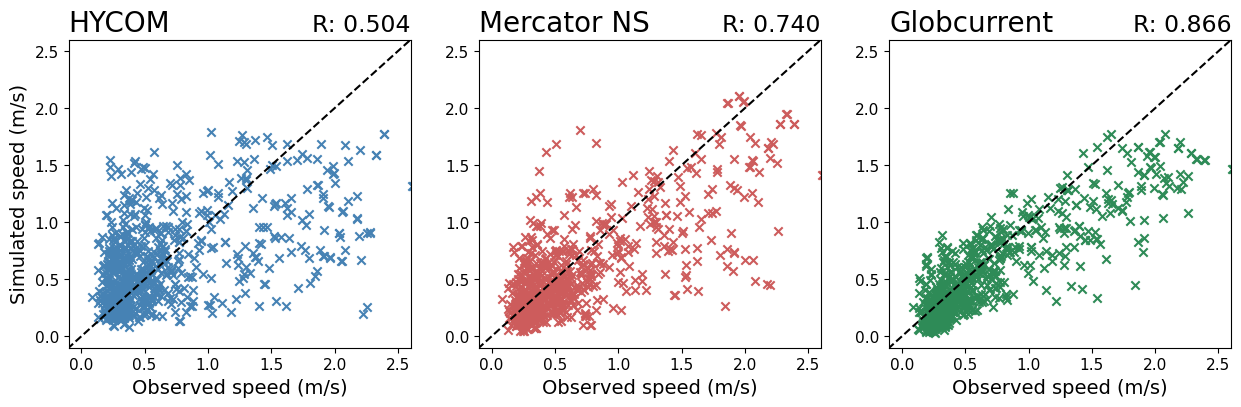}
    \caption{Scatter plots of the mean observed drifter speed per 24h segment against the mean simulated drifter speed per 24h simulation for the three different ocean surface current products, HYCOM, Mercator NS and Globcurrent, with correlation coefficient (R). The black, dashed line represents the 1:1 ratio.}
    \label{fig:ocean_models_speed}
\end{figure*}

In figure \ref{fig:ocean_models_speed} the mean simulated drifter speeds of every 24h simulation are compared to the mean observed drifter speeds of the same 24h windows. 
The correlation coefficients for HYCOM, Mercator NS and Globcurrent are 0.504, 0.740 and 0.866 respectively.  
Although Globcurrent shows the highest correlation in speed, it underestimates for high velocities.
Globcurrent never provided velocities higher than 1.8 m/s while the observed drifter velocities in the Agulhas Current are seen to be between 2.0-2.5 m/s.
This underprediction of strong currents by the Globcurrent product in the Agulhas System is observed in earlier research by \cite{Hart-Davis2018} and \cite{Barnes2020}.
Due to the relatively coarse grid of the Globcurrent product, a strong, narrow current as the Agulhas Current gets smoothed (\cite{Quilfen2018}). 
HYCOM also underestimates the largest current speeds, with drifter speeds never over 1.8 m/s, but also overestimates the speed when the observed drifter speeds are low.
The Mercator NS model shows a fairly good correlation with the observed speeds (R = 0.740) and is able to simulate higher drifter velocities (up to 2.2 m/s). 
Although Mercator NS shows larger velocities for high current areas, it still underestimates for a large part of the strongest currents. 

To compare the two model based products, the Improvement Factor (IF) is calculated compared to the, observation based, Globcurrent simulations (Eq. \ref{eq:IF}). 
The Lagrangian simulation performance using HYCOM is 39.7\% and 37.9\% lower than with the Globcurrent product for LWS and MCSD respectively (Table \ref{tab:ocean_forcing}).
The Mercator NS performance is 14.6\% lower than Globcurrent for LWS and 14.0\% for MCSD.
The percentile improvement in LWS and MCSD show similar values. 
This strengthens that the ocean model performance results are valid.

\begin{table}[]
    \centering
    \begin{tabular}{m{2.2cm} m{1.6cm} m{1.8cm} m{1.1cm}}
         \hline
         Ocean forcing & LWS (\%) & MCSD (\%) & R speed \\ \hline
         HYCOM & -39.7 & -37.9 & 0.50 \\
         Mercator NS & -14.6 & -14.0 & 0.74 \\
         Mercator Tot. & -9.6 & -10.6 & 0.76 \\ 
         Globcurrent &  &  & 0.87 \\ \hline
    \end{tabular}
    \caption{Improvement factor (Eq. \ref{eq:IF}) of the performance of simulations with different ocean model forcing compared to simulations with Globcurrent ocean forcing (observation-based) and the correlation coefficient (R) between observed and simulated drifter speed.}
    \label{tab:ocean_forcing}
\end{table}

\subsection{Additional forcing}
\begin{figure*}
    \centering
    \includegraphics[width=\linewidth]{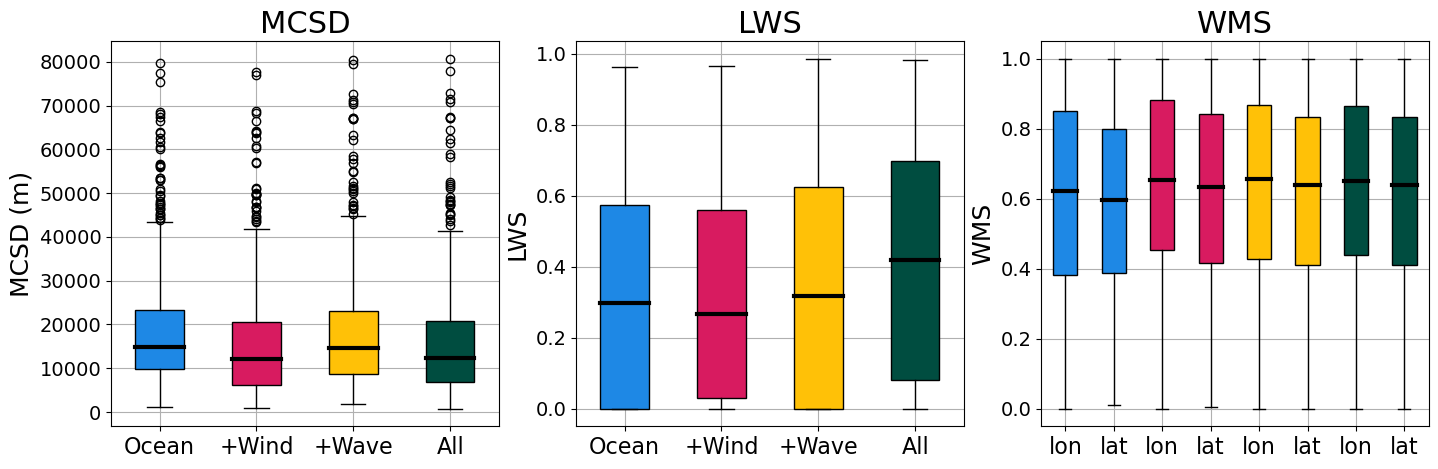}
    \caption{Box and whisker plots for MCSD (Eq. \ref{eq:mcsd}), LWS (Eq. \ref{eq:liuweis}) and WMS for longitude and latitude (Eq. \ref{eq:liuweis}) between the six observed drifter trajectories and the OpenDrift simulations with different forcing set-ups. The box extends from the lower to upper quartile of the data and the horizontal line within the box represents the median value. The whiskers are plot from the lower quartile of data minus 1,5 times the IQR to the upper quartile plus 1,5 times the IQR. Flier points are plotted for data points outside the whisker range.}
    \label{fig:general_metrices}
\end{figure*}

Four experiments, of each 707 simulations, are conducted with different forcing set-ups including/excluding additional wind and/or wave forcing (Table \ref{tab:sims}, exp. 5-8). 
In figure \ref{fig:general_metrices} the range and median of the MCSD, LWS and WMS for longitude and latitude are displayed for four different experiments. 
Mainly, the addition of the wind field results in smaller MCSD compared to only ocean forcing. 
However, looking at the LWS, the addition of the wave forcing results in higher skill scores than only ocean and ocean plus wind forcing.
Nevertheless, the addition of both wind and waves results in the highest skill scores. 
For the WMS, the differences between the different forcing set-ups are less pronounced. 
The largest increase in WMS, compared to only ocean forcing, is observed for the latitudinal direction. 
Besides, a difference in skill is observed between latitudinal and longitudinal direction. 
This difference between the directions decreases the most by the addition of wind forcing. 
This could be explained by a dominant wind direction.
With a predominant wind direction from the North and/or the South during the period of the experiment, the wind-induced drift (direct wind drag, stress and Stokes drift) of the buoys is larger in latitudinal direction. 
So to say, the wind-induced drift makes up a larger part of the total drift in latitudinal then longitudinal direction. 
Without additional wind and/or wave forcing, this wind-induced drift is not well resolved, resulting in a lower simulation skill for the latitudinal direction. 

Additionally, the Improvement Factor (Eq. \ref{eq:IF}) between the experiment with only ocean forcing and the experiments with additional wind and/or wave forcing is calculated (table \ref{tab:results_if}). 
The simulations with ocean, wind and wave forcing show a 23\% increase in LWS and a 16\% improvement of the MCSD. 
For the addition of, solely, wind forcing, contradictory results are shown in MCSD (+19\%) and the LWS (-4\%). 

Finally, the speed. 
The correlation coefficient (R) between the simulated and observed drifter speeds can be found in table \ref{tab:conditions_speed} in the 'All data' column. 
The addition of extra forcing fields leads to an increase in R from 0.74 for only ocean forcing to 0.81, 0.77 and 0.82, for the addition or wind, waves and both, respectively.

\begin{table}[]
    \centering
    \begin{tabular}{m{1.6cm} m{1.6cm} m{1.6cm} m{1.6cm}}
         \hline
         Forcing & LWS (\%) & MCSD (\%) & R speed \\ \hline
         +Wind & -4 & +19 & 0.81 \\
         +Wave & +7 & +5 & 0.77 \\
         All & +23 & +16 & 0.82 \\ \hline
    \end{tabular}
    \caption{The Improvement Factor (IF, Eq. \ref{eq:IF}) of simulations with wind, wave or wind and wave (all) forcing added in comparison to only ocean forcing (Mercator NS) and the correlation coefficient (R) between observed and simulated drifter speed.}
    \label{tab:results_if}
\end{table}

\subsubsection{Wind Drift Factor}
For the simulations including wind forcing, per segment, 25 simulations have been run with a WDF ranging from 0.00-0.05. 
For every segment, the WDF of the simulation with the highest WMS, in longitude and latitude, is obtained. 
This optimal WDF for the OMB is found to be $\sim 0.018$ and $\sim 0.023$ for the simulations including and excluding additional wave forcing (table \ref{tab:general_wdf}). 

The obtained optimal WDF of the OMB is compared to those of the LainePoiss-v0.1 (LP) and iSphere drifters in earlier research. 
The LP and iSphere drifters are undrogued and similar in size to the OMB, but have a rounded shape that sticks out from the ocean surface.
The OMB is designed to be up to 80\% submerged in the ocean surface and is therefore expected to be less affected by the wind than the LP and iSphere drifters. 
\cite{Part2023} found a WDF of 2\% for the LP drifters including ocean, wind and Stokes drift forcing, and 3\% excluding Stokes drift. 
For the iSphere drifters, \cite{Sutherland2020} reported WDFs of 1.6\% and 2.8\% and \cite{Dagestad2019} founded 3\% and 4\%, respectively. 
Besides, \cite{Jones2016} and \cite{Brekke2021} both reported a WDF of 2\% for the iSphere drifter, including ocean, wind and Stokes drift. 
The OMB, thus shows a lower to comparable optimal WDF compared to the iSphere and LP drifters, as expected.  
There must be noted that the optimal WDF shows a high seasonal and regional dependency due to mixed layer stratification (\cite{Rio2003}) and the rate of change in the wind forcing (\cite{Rohrs2015}). 
One should be aware of this, using an optimal wind drift factor obtained by other research at different location and time. 

For the simulations including the Stokes drift forcing, a lower mean optimal WDF is found (0.017/0.018) than for simulations without the Stokes drift forcing (0.021/0.024). 
Locally generated wind waves will create a Stokes drift which is aligned, in direction and power, to the wind field. 
If there is no Stokes drift forcing used for the Lagrangian simulation, the empirically derived optimal WDF will, besides wind drag, also cover for the Stokes drift of the local wind waves (and swell waves, if their direction is aligned with the local wind direction).
Therefore, an empirically derived WDF is smaller for simulation including wave forcing then excluding wave forcing.

Comparing the optimal WDFs of the simulations with and without Stokes drift forcing can give an indication for the relative impact of the direct wind drag and Stokes drift on the OMB during this experiment. 
For this, it's assumed that all the Stokes drift is from local wind waves and thus in the same direction as the wind.
This is not a very good assumption for this region as it is subject to Southwesterly swell (\cite{Rautenbach2020}), but it will give a first indication.
From table \ref{tab:general_wdf}, one can calculate that the Stokes drift is as large as ~29\% of the direct wind drag. 
This indicates that the direct wind drag is a (3.5 times) larger contributor to the drift of the OMB than the Stokes drift of the local wind waves.

\begin{table}[]
    \centering
    \begin{tabular}{m{2.7cm} m{2.3cm} m{2.3cm}}
         \hline 
         \textit{Optimal WDF} & Lon (range) & Lat (range)  \\ \hline
         Ocean+Wind & 0.024(0.00-0.05) & 0.021(0.00-0.05) \\
         Ocean+Wind+Wave & 0.018(0.00-0.05) & 0.017(0.00-0.05) \\ \hline
    \end{tabular}
    \caption{The mean and range of the optimal wind drift factor based on the WMS in longitude and latitude.}
    \label{tab:general_wdf}
\end{table}

\subsection{Conditional Analysis} \label{sec:results_conditions}
In this section, the simulations will be analyzed for extreme conditions. 
This includes, ocean current > 1.25 m/s, wind speed > 10 m/s and Stokes drift > 0.15 m/s. 
In figure \ref{fig:conditions_lw}, the LWS is plotted for all the data and the extreme conditions per forcing set-up. 

\subsubsection{Ocean Current}
\begin{figure}
    \centering
    \includegraphics[width=\linewidth]{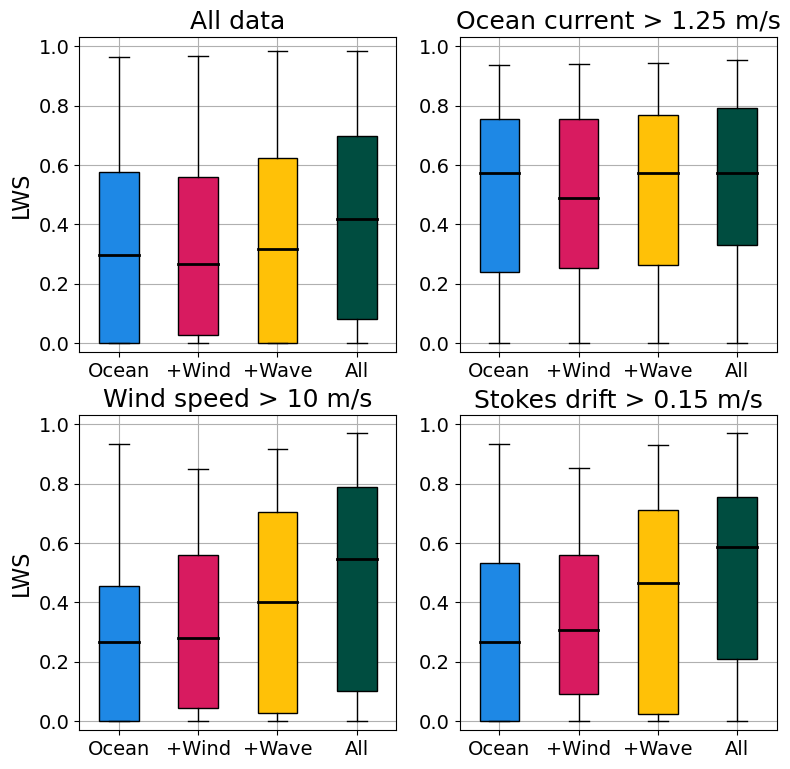}
    \caption{Box and whisker plots for the LWS (Eq. \ref{eq:liuweis}) between the six observed drifter trajectories and the OpenDrift simulations with four different forcing set-ups. The results are shown for extreme ocean (>1.25 m/s), wind (>10 m/s) and wave (>0.15 m/s) conditions. The box extends from the lower to upper quartile of the data and the horizontal line within the box represents the median value. The whiskers are plot from the lower quartile of data minus 1,5 times the IQR to the upper quartile plus 1,5 times the IQR. Flier points are plotted for data points outside the whisker range.}
    \label{fig:conditions_lw}
\end{figure}

For large ocean current, the observed skill scores are in general higher compared to the total dataset ('All data' in fig.\ref{fig:conditions_lw}).
These high skill scores are expected because the largest ocean currents are found in the Agulhas Current. 
The Agulhas Current is a very stable, strongly baroclinic flow, due to the very steep continental slope along its path. 
Therefore, ocean models tend to resolve the location of the Agulhas Current quite well and the simulated paths are expected to be in good alignment with the observed drifter trajectories.
However, the strong currents in the Agulhas Current are underpredicted by the Eulerian ocean models (as discussed in the ocean model results, section \ref{sec:results_ocean_forcing}, fig. \ref{fig:ocean_models_speed}), 
Besides, the addition of wind and wave forcing results in less improvement of the LWS when there is a large ocean current regime than for the complete dataset.
The use of all forcing fields resulted in an improvement factor of only 11\% for an ocean current > 1.25 m/s, instead of the 23\% increase for all the data together. 
This is expected, as when the ocean current is larger, the Stokes drift and direct wind drag are making up a relatively smaller portion of the total surface drift, which makes their effect smaller. 

\subsubsection{Wind \& Waves}
For large wind speed and Stokes drift, another pattern is observed. 
The skill scores for the simulations, strongly increase by addition of the wind and the wave field as a forcing. 
For addition of wind, wave and both fields, we find an improvement factor in LWS of about 7,5\%, 33\% and 58\%, respectively. 
For all the data together, an improvement factor for addition of wind, wave and both fields is found to be -4\%, 7\% and 23\%, respectively. 
This shows that the addition of wind and/or wave forcing in Lagrangian trajectory simulations is of larger importance for the model performance during periods of high wind speed and/or the Stokes drift. 
This is expected, because for high Stokes drift and/or wind speed, the surface drift caused by those components will make up a relatively larger part of the total drift of the object, then for low wind and/or wave conditions. 

\subsubsection{Wind Drift Factor}
\begin{figure}
    \centering
    \includegraphics[width=\linewidth]{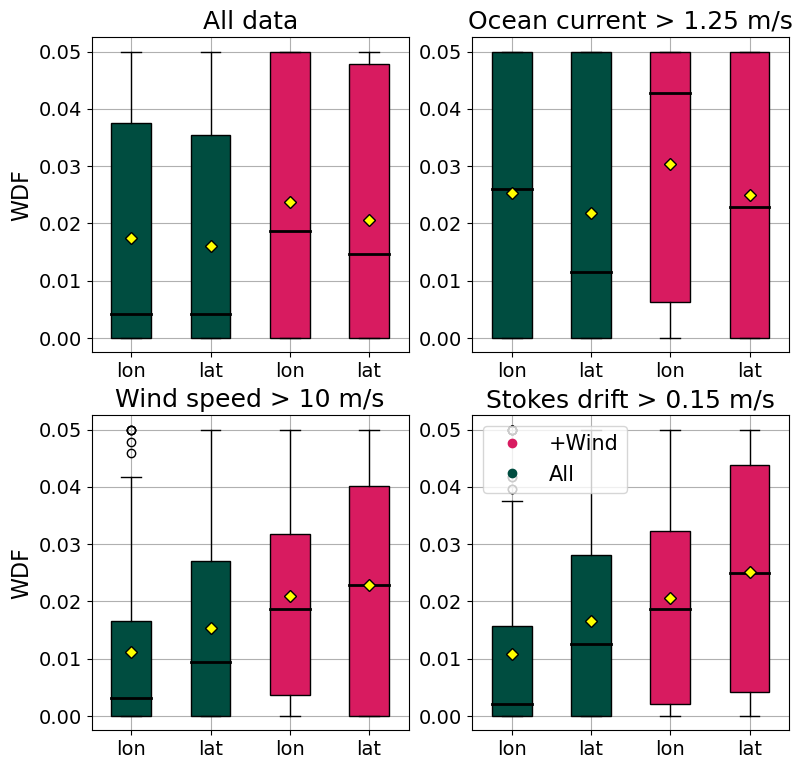}
    \caption{Box and whisker plots for the wind drift factor that resulted in the simulation with the highest WMS (Eq. \ref{eq:liuweis}), in longitude and latitude separately. The results are shown for all the data and extreme ocean (>1.25 m/s), wind (>10 m/s) and wave (>0.15 m/s) conditions. The box extends from the lower to upper quartile of the data and the horizontal line within the box represents the median value. The whiskers are plot from the lower quartile of data minus 1,5 times the IQR to the upper quartile plus 1,5 times the IQR. Flier points are plotted for data points outside the whisker range.}
    \label{fig:conditions_wdf}
\end{figure}

Furthermore, the optimal WDF for the different conditions is analysed. 
In figure \ref{fig:conditions_wdf} the optimal WDFs, found by WMS, for different forcing set-ups are plotted for all data together and for large ocean current, wind speed and Stokes drift. 
For large ocean current regimes, the mean optimal WDF is found to be higher than for all data, in both forcing set-ups.
Besides, a large spread in optimal WDFs is observed with more simulations showing an optimal WDF of 5\% then for all data.
The Eulerian ocean models underestimate for large ocean currents (Section \ref{sec:results_ocean_forcing}). 
If the wind direction is aligned with the current direction, a higher WDF will result in higher drifter speeds.
In this way, the WDF can compensate for the underestimation of ocean current by the ocean model forcing in the Agulhas Current. 
In addition, it is observed that when the ocean current is this large, the simulated trajectories with different WDFs do not show large differences in their path as the ocean current much larger and therefore dominant.
The empirically derived, larger, mean optimal WDF is therefore not expected and seems to be correlated to the underprediction of the ocean current instead of wind processes. 

For the high wind speed and large Stokes drift regimes, the optimal WDF results are similar and will be discussed as one. 
Compared to all the data, the high wind and Stokes regimes show a smaller spread in optimal WDFs with a lower mean optimal WDF in longitudinal direction than latitudinal direction. 
The mean optimal WDFs for the simulations with ocean, wind and Stokes forcing are found to be lower (1.1-1.5\%) than for the total dataset (1.7-1.8\%). 
An explanation for this could be found if the wind and Stokes drift direction are aligned.
In this case, the Stokes drift becomes more important than the direct wind drag when the Stokes drift is large. 
This is also seen by \cite{Parn2023} in the Baltic Sea.
They found that for large significant wave height and alignment of the wind and wave direction, the Stokes drift becomes more important than the wind drag. 
In the research by \cite{Haza2019} on undrogued drifters in the Northern Gulf of Mexico, higher optimal windage values were found for low winds (3-3.5\%) than for high winds (2-3\%). 
\cite{Haza2019} explain this by a larger floater slip for low wind conditions.
This means that, during low wind conditions, the drifter will move more freely from the water surface instead of with ('attached to') the water surface. \\
\\
\subsubsection{Speed}
Lastly, the correlation between simulated and observed drifter speed (table \ref{tab:conditions_speed}) will be analysed.
For large ocean currents, the correlation coefficient between the simulated and observed drifter speed is found to be lower compared to R of the total dataset for all forcing set-ups. 
This is high likely due to the underestimation of large ocean currents by the Mercator NS ocean model, as seen in figure \ref{fig:ocean_models_speed}.
For high wind speeds, R between simulated and observed speeds is found to be a little lower or similar to the complete dataset, except for the simulation with only ocean model forcing. 
R is significantly lower for the simulations with only ocean forcing for high wind speeds. 
This is logical, as when the wind speed is high, the direct wind drag will be a relative larger contribution to the total surface drift compared to when the wind speed is low. 
Therefore, the addition of a wind velocity field is of increasing importance when the wind speeds are larger to obtain realistic surface drift velocities. 
Contrary to the large ocean current and high wind speed conditions, large Stokes drift shows an increase in R between observed and simulated drifter speeds compared to all the drifter data. 
The largest increase in R is found for the simulation where wave forcing is included, stressing that the addition of the wave field is more important to obtain realistic drifter speeds in the simulation when the Stokes drift is large.
Although, the addition of a wind field shows a similar R as the addition of the wave field. 
The addition of the wind field can already partly cover for the larger Stokes drift if the waves are locally generated wind waves and not elsewhere generated swell waves. 

\begin{table*}[]
    \centering
    \begin{tabular}{m{2.6cm} m{3.6cm} m{3.6cm} m{3.6cm} m{1.2cm}}
        \hline
        \textit{R speed} & Ocean current > 1.25 m/s & Wind speed > 10 m/s & Stokes drift > 0.15 m/s & All data  \\ \hline
        Ocean forcing & 0.61 & 0.58 & 0.72 & 0.74 \\
        + wind forcing & 0.70 & 0.77 & 0.84 & 0.81 \\
        + wave forcing & 0.61 & 0.75 & 0.83 & 0.77 \\
        All forcing & 0.69 & 0.81 & 0.89 & 0.82 \\ \hline
    \end{tabular}
    \caption{Correlation coefficient between the simulated and observed drifter speed for different forcing set-ups for all data and 'extreme' conditions in ocean current, wind speed and Stokes drift.}
    \label{tab:conditions_speed}
\end{table*}

\section{Discussion \& Conclusion}\label{Discussion}
To assess the performance of the Lagrangian trajectory simulations, three different aspects have been analysed: (1) two different offline Lagrangian tracking tools, OpenDrift and Parcels, (2) three different Eulerian ocean current products, HYCOM, Mercator and Globcurrent, and (3) the addition of wind and/or wave forcing parameterizations. 
Moreover, the impact of 'extreme' conditions of ocean current, wind speed and Stokes drift forcing in Lagrangian trajectory simulations has been evaluated. 

\subsection{Lagrangian simulation models}
We can conclude that with the same time-stepping scheme and linear interpolation methods, the Lagrangian trajectory models OpenDrift and Parcels perform identical.
When the ocean current is strong, the interpolated ocean current by the two models tend to show a small deviation from each other, but overall the two models show a perfect correlation in speed, MCSD and LWS compared to the observed drifter trajectories.
Lagrangian simulation models differ more from each other in their features, as user-friendliness, efficiency of data in- and output, offline/online calculations, interpolation possibilities, etc. (\cite{VanSebille2018}) 
The choice of Lagrangian simulation models needs to be considered in regards to these conditions.

\subsection{Ocean Models}
By comparing the Eulerian ocean models (HYCOM and Mercator) and the altimeter-based Globcurrent fields, large differences in performance are observed. 
The observation-based Globcurrent product shows larger skill, lower MCSD and higher correlation in speed with the observed drifter trajectories.
This is in agreement with the results from \cite{Hart-Davis2018}, where simulations with Globcurrent were found to be 25\% more accurate than the simulations with NCOM (The Navy Coastal Ocean Model) ran by \cite{Barron2007} in the Greater Agulhas System. 
This confirms the hypothesis that the ocean currents derived by the Globcurrent product are more accurate although at coarser resolution. 
The HYCOM model performance is about 39\% lower than the Globcurrent product and shows the lowest correlation in drifter speed (0.504).
This is an interesting result as the HYCOM model is one of the default ocean forcing models in OpenDrift. 
The simulations with the Mercator NS model component show 14\% lower accuracy than the Globcurrent simulations, suggesting that Mercator NS ocean currents are more accurate than the currents from the HYCOM model. 
Although Globcurrent showed the highest skill, it underestimates the speed for strong ocean currents due to smoothing of the velocity field in strong and narrow currents (\cite{Quilfen2018}); \cite{Hart-Davis2018}).
This is further supported by \cite{Marzocchi2015}, who demonstrated an improvement in the velocities of the Northern Atlantic surface current through an increase in the resolution of the input model. 
In comparison, the Mercator NS model is able to predict higher velocities than the Globcurrent product, which is also reported by \cite{Barnes2020}.

\subsection{Stokes drift and WDF parameterization}
Finally, the use of additional wind and/or wave velocity forcing in Lagrangian trajectory simulations was examined.
When including both WDF parameterization and Stokes drift forcing, resulted in the highest performance compared to the observed drifter trajectories with an increase in skill score (+23\%), separation distance (+16\%) and speed compared to trajectory simulations with only ocean forcing.

\subsubsection{Wind Drift Factor}
A large spread is observed in the choice of WDF with an optimal WDF of 0 for more than one third of the simulated trajectories. 
This spread is even larger for stronger ocean current regimes, suggesting that for these conditions and drifters, the Eulerian ocean current is the main driver of the drift while the wind parameterization appears unclear. 
This motivates a discussion on the ambiguity of the WDF parameterization in Lagrangian trajectory simulations. 

When the WDF is empirically derived, it accounts for many processes as windage, wave-induced drift, unresolved ocean current vertical shear (\cite{Tamtare2021}; \cite{Laxague2018}) and other dynamics that are not directly related to the wind  (\cite{Jones2016}). 
Moreover, the optimal WDF is highly dependent on factors such as the region (\cite{Rio2003}; \cite{Poulain2009}), the season (\cite{Rio2003}) and the temporal rate of change of the wind field (\cite{Rohrs2015};\cite{DeDominicis2016}) as well as the object of interest (\cite{Hart-Davis2018}). 

In consistence with these findings we achieved significantly improved correlation between observed and simulated drifter speed and a larger mean optimal WDF when the wind field forcing was included.
It is also revealed that the Eulerian ocean models tend to underestimate strong ocean current (Figure \ref{fig:ocean_models_speed}). 
All in all, it therefore seems that the optimized WDF is not just solving for the direct wind drag, but is also compensating for the underestimation of the Eulerian ocean current.

Preferably, we should have a more theoretical basis on the forces that act on a floating object at the ocean surface, instead of using empirical parameterizations as the WDF, or all other, not wind-related, processes should be better resolved in the global ocean models.
An example of a more physical based solution for wind related processes at the ocean surface is introduced by \cite{Tamtare2021}.
They argue that the WDF does not properly invoke the vertical shear of the horizontal current in the upper ocean, that makes up for a great part of the wind-related processes that contribute to the surface drift.  
To incorporate the effect of the vertical shear in a more realistic way, \cite{Tamtare2021} made use of an extrapolated ocean current.
Their simulations showed higher skill and the empirically derived WDF was lower and more related to the direct wind drag only.
An approach like this could therefore be a solution towards a more physically based incorporation of direct wind drag and would be interesting for further research.

Contrary to discussed above, multiple drifter studies report most accurate results for Lagrangian simulations when incorporating only wind forcing (e.g., \cite{Haza2019} for low wind conditions; \cite{Paldor2004}; \cite{Dagestad2019}). 
This is explained by the higher accuracy of atmospheric models than ocean models, due to the much larger scales and better coverage of atmospheric observations.
Despite substantial advancements in assimilating satellite data, the complex and non-linear character of advection in the ocean remains a challenge in ocean modelling.
A skillful use of the wind could, therefore, potentially reduce this error for the near-surface ocean current.
\cite{Haza2019} and \cite{hole2019deepwater} found that under low wind conditions, the highest accuracy is obtained for the wind-forced only simulations (Northern Gulf of Mexico).
This is in contrast to the results reported in this paper where we have studied very areas with very strong currents that are not wind driven. 
During low (high) wind conditions, the addition of wind forcing reduces (increases) the simulation performance. 
It is seen that resolving dominant dynamic structures in the ocean current, like the location of the Retroflection loop and major eddies, is of great importance for the performance of Lagrangian trajectory simulations in the Agulhas Current System.
Hence, the Eulerian ocean current forcing is the dominant factor for the ocean surface drift of the OMBs in the Agulhas Current System.

\subsubsection{Waves}
It is common practice in Lagrangian modelling to account for the Stokes drift in the WDF under the assumption that the wind speed and the Stokes drift are linearly correlated (\cite{Breivik2008}). 
This study shows, that this is not a valid assumption for the OMBs in the Agulhas Current System.
In many cases, uncertainties are clearly reduced when invoking the Stokes drift forcing from a numerical wave model into the Lagrangian trajectory simulations. 
This has several explanations. 
To start, the wave height of locally-generated wind waves depends not only on the local wind speed, but also on the duration and the fetch of the wind field. 
Moreover, a young wind-wave field is characterized by short waves with a large Stokes drift and vertical shear, while for swell wave fields, the Stokes drift is more uniformly distributed with depth (\cite{Rohrs2012}). 
In addition, the forecast skill of an ocean model is usually lower than the forecast skill of a numerical wave model as it is usually on par with the skill of the atmospheric forecast model that forces the wave model (\cite{Rohrs2012}).
However, a decrease in skill is reported in some studies when the Stokes drift forcing is included in the Lagrangian trajectory simulations due to the fact that the predictability of the Stokes drift through a wave model is inferior to the direct predictability of the wind (e.g.\cite{Dagestad2019} at Norwegian coast).

In this research, the addition of Stokes drift forcing to the Lagrangian simulations shows a significant increase in performance, especially for high Stokes drift regimes (>0.15 m/s).
Including Stokes drift forcing to the simulations, is therefore even more important when the Stokes drift is large, as it will make up a larger part of the total ocean surface current.
This is also supported by the results of the optimal WDF. 
A higher optimal WDF is obtained for low wind conditions than high wind conditions, like \cite{Haza2019} found for undrogued drifters in the Northern Gulf of Mexico.
A larger floater slip during low wind conditions will cause the drifter to move more freely from the water surface, instead of 'attached to' the water surface (\cite{Haza2019}).
Likewise, \cite{Parn2023} found that the Stokes drift is of higher importance than wind drag, in case of large significant wave height and aligned wind and wave direction (Baltic Sea).
All above, supports the conclusion that the addition of Stokes drift forcing is important to obtain simulations closer to observations, and the Stokes drift is an important factor in the surface ocean drift in the Agulhas Current System.

However, the addition of Stokes drift forcing is still a parameterization of the wave effects on the ocean surface current.
Other wave related processes like the flux of momentum and turbulent kinetic energy from the waves to the ocean from wave dissipation (through wave breaking), remain unresolved.
Without accounting for the wave-induced fluxes, simply adding Stokes drift to the Eulerian current, contradicts with the principle of momentum conservation and has a negative effect on the ocean current predictions (\cite{Rohrs2012}).
Therefore, \cite{Rohrs2012} suggest that the wave-induced fluxes of momentum energy and the Coriolis-Stokes force should be accounted for in the computation of the Eulerian ocean current to better resolve the wave-related processes.
Another way to cover for more wave related processes, is shown by the coupling of a global ocean model into the wave forecast (\cite{Barnes2020}). 
In addition, wave-current interactions can also be of great importance. 
\cite{Barnes2020} showed that the ocean current strongly affects the significant wave height when the wave propagation direction opposes or follows the current in the Agulhas Current System. 
To further improve Lagrangian simulations, it is therefore necessary to advance the account of wave-related effects and wave-current interactions. \\
\\
Overall, the performance of the Eulerian ocean model seems to be the most important aspect in the skill of the Lagrangian trajectory simulations in the Agulhas Current region.
However, although the addition of wind and wave parameterizations improves the Lagrangian simulation performance, other, more physical-based parameterizations or methods are needed to truly resolve the ocean surface current. 
 
\subsection{Data limitations}
A final remark should be made regarding the limited scope of the drifter dataset used in this study.
For the validation of an ocean model using Lagrangian trajectory simulations and observed drifter trajectories, it is advantageous to compare with a substantial number of observed drifter trajectories. 
A larger drifter dataset ensures a broader capture of various ocean processes and dynamics with diverse spatial and temporal scales and variability.
Six drifter trajectories were used in this study over a period of two months. 
In comparison, \cite{Hart-Davis2018} made use of 1041 drifters from the global drifter programme (GDP) deployed in the Agulhas Current System in the period 1993-2015 to assess the Globcurrent products.   \\
\\

% Uncomment and use as the case may be
%\begin{theorem} 
%\end{theorem}

% Uncomment and use as the case may be
%\begin{lemma} 
%\end{lemma}


\textbf{Acknowledgements} \\
\\
This work is a part of the One Ocean Expedition, which involves Stiftelsen Seilskipet Statsraad Lehmkuhl and several scientific institutions: The Norwegian Institute of Marine Research, The University of Bergen, The Norwegian Research Center, The Norwegian Meteorological Institute, Western Norway University of Applied Science, Nansen Environmental and Remote Sensing Center, and The Norwegian Institute for Water Research. 
The scientific work during One Ocean Expedition was sponsored by the Agenda Vestland / Sparebanken Vest Foundation. 

This study was funded by the Agenda Vestland / Sparebanken Vest Foundation, the Norwegian Research Council and the European Space Agency (ESA). 
The deployment of the surface wave drifting buoys and the associated data collection was made possible by the One Ocean Expedition onboard the tall ship Statsraad Lehmkuhl during the science and sail voyage from Maputo (Mozambique) to Cape Town (South Africa). 
This leg was coordinated and led by the Nansen Environmental and Remote Sensing Center together with ESA and OceanDataLab. 
In addition the voyage saw participation of several research institutes and universities including: The Norwegian Institute of Marine Research, The University of Bergen, The Norwegian Meteorological Institute, Western Norway, University of Applied Science, The Norwegian Institute for Water Research and the OceanDataLab (Plouzane, France). 

%\section{Appendix A}\label{}

\printcredits

\bibliographystyle{cas-model2-names}

\bibliography{cas-refs}

\bio{}
\endbio

\end{document}